\documentclass[11pt]{article}
\pdfoutput=1

\usepackage{epsfig}
\usepackage{graphicx}
\usepackage{epstopdf}
\usepackage{float}

\usepackage{amsfonts}

\usepackage{amsmath}                                                    
\usepackage{amsthm}                                                     
\usepackage{amssymb}
\usepackage{multirow}
\numberwithin{equation}{section} %section #s
\newcommand{\newc}{\newcommand}
\newc{\nn}{\nonumber}
\newc{\be}{\begin{equation}}
\newc{\ee}{\end{equation}}
\newc{\ba}{\begin{eqnarray}}
\newc{\ea}{\end{eqnarray}}
\newc{\ov}{\overline}
\newc{\bg}{\begin{gathered}}
\newc{\eg}{\end{gathered}}
\newc{\tref}[1]{Table \ref{#1}}
\newc{\eref}[1]{Equation \eqref{#1}}
\newc{\su}[1]{$SU(#1)$}
\newc{\bm}[1]{\mathbf{#1}}
\newc{\fref}[1]{Figure \ref{#1}}

\begin{document}
\begin{titlepage}

%\vspace*{-15mm}
%\begin{flushright}
%SHEP-11-XX\\
%\end{flushright}
\vspace*{0.7cm}

\begin{center}
{
\bf\LARGE 750 GeV 
Diphoton excess  from $E_6$ in F-theory GUTs}
\\[12mm]
Athanasios~Karozas$^{\dagger}$
\footnote{E-mail: \texttt{akarozas@cc.uoi.gr}},
Stephen~F.~King$^{\star}$
\footnote{E-mail: \texttt{king@soton.ac.uk}},
George~K.~Leontaris$^{\dagger}$
\footnote{E-mail: \texttt{leonta@uoi.gr}},
Andrew~K.~Meadowcroft$^{\star}$
\footnote{E-mail: \texttt{a.meadowcroft@soton.ac.uk}}
\\[-2mm]

\end{center}
\vspace*{0.50cm}
\centerline{$^{\star}$ \it
Physics and Astronomy, University of Southampton,}
\centerline{\it
SO17 1BJ Southampton, United Kingdom }
\vspace*{0.2cm}
\centerline{$^{\dagger}$ \it
Physics Department, Theory Division, Ioannina University,}
\centerline{\it
GR-45110 Ioannina, Greece}
\vspace*{1.20cm}

\begin{abstract}
\noindent
We interpret the 750-760 GeV diphoton resonance as one or more of the spinless 
components of a singlet superfield arising from the three 27-dimensional representations
of $E_6$ in F-theory, 
which also contain three copies of colour-triplet charge $\mp 1/3$ vector-like
fermions $D_i,\bar{D}_i$ and inert Higgs doublets to which the singlets may couple.
For definiteness we consider (without change) a model that was proposed some time ago
which contains such states, as well as bulk exotics, leading to gauge coupling unification. 
The smoking gun prediction of the model is the existence of other similar spinless
resonances, possibly close in mass to 750-760 GeV,
decaying into diphotons, as well as the three families of vector-like fermions
$D_i,\bar{D}_i$.
\end{abstract}

\end{titlepage}

\thispagestyle{empty}
\vfill
\newpage

%%%%%%%%%%%%%%%%%%%%%%%%%%%      START     

\section{Introduction }

Recently ATLAS and CMS experiments have reported  an excess of 14 and 10 diphoton events at an invariant mass 
around 750 GeV and 760 GeV from gathering data at LHC Run-II with $pp$ collisions at the center of mass energy of 13 TeV \cite{ATLAS,CMS}.
The  local significance of the ATLAS events is 3.9 $\sigma$ while that of the 
CMS events is  2.6 $\sigma$, corresponding to cross sections $\sigma(pp\to\gamma\gamma)=10.6$ fb
 and $\sigma(pp\to \gamma\gamma)=6.3$ fb.
ATLAS favours a width of $\Gamma \sim 45$ GeV, while CMS, while not excluding 
such a broad resonance, prefers a narrow width.
The Landau-Yang theorem implies spin 0 or 2 are the only possibilites for 
a resonance decaying into two photons.
The only modest diphoton excesses observed by ATLAS and CMS at this mass scale
may be (at least partially) understood by the factor of 5 gain in cross-section due to gluon production. However there is no evidence for any coupling of the resonance into anything except gluons and photons (no final states such as $t\bar{t}$, $b\bar{b}$, $l\bar{l}$, $ZZ$, 
$WW$, etc., with missing $E_T$ or jets have been observed).

If these facts are confirmed by future data, 
it will be the first indication for new physics at the TeV scale and possibly
a harbinger of more exciting discoveries in the future.  These findings also pose 
a challenging task for theoretical extensions of the Standard Model (SM) spectrum. Several
interpretations have  been suggested based on extensions of the Standard Model spectrum ~\cite{Harigaya:2015ezk}-\cite{Ibanez:2015uok}.
Many of these papers suggest a spinless singlet coupled to vector-like fermions
\cite{Harigaya:2015ezk,Pilaftsis:2015ycr,Franceschini:2015kwy,Ellis:2015oso,Gupta:2015zzs,
Cao:2015pto,Kobakhidze:2015ldh,Falkowski:2015swt,Kim:2015ron,
Ding:2015rxx,Heckman:2015kqk,Kim:2015ksf,Cvetic:2015vit,Allanach:2015ixl,Wang:2015omi,Cai:2015hzc,Anchordoqui:2015jxc,Ibanez:2015uok}.
Indeed, the observed resonance could be interpreted as a Standard Model scalar or pseudoscalar
singlet state $X$  with mass $m_X\sim 750-760$ GeV. 
The process of generating the two photons can take place by the gluon-gluon fusion mechanism
according to the process 
\[ gg \to X\to \gamma\gamma \]
hence it requires production and decay of the particle $X$. In a renormalisable theory 
this interaction  can be  realised assuming vector-like multiplets $f+\bar f$  at the TeV scale,
where $f$ carry electric charge and colour. 
Such vector like pairs have not been observed at LHC, hence the mass of the fermion pair 
$M_f$ is expected roughly to be at or above the TeV scale, $M_f\gtrsim 1$ TeV. 

If this theoretical interpretation is adopted, effective field theory models derived in the context 
of String Theory are excellent candidates to accommodate the required  states.  
Indeed, singlet scalar fields are the most common characteristic of String Theory effective models.
These can be either scalar components of supermultiplets or of pseudoscalar nature such as 
axion fields having direct couplings to gluons and photons and therefore relevant to the observed process. However another aspect of string theory interests us here,
namely that in the low energy spectrum of a wide class of string models vector-like supemultiplets either 
with the quantum numbers of ordinary matter or with exotic charges are generically
present. Moreover, 
in specific constructions they can remain in the low energy spectrum and get a mass at the TeV scale. A particularly elegant possibility is that the low energy spectrum consists of 
the matter content of three complete
27-dimensional representations of $E_6$, as in the E$_6$SSM 
\cite{King:2005jy}, or minimal E$_6$SSM \cite{Howl:2007zi},
minus the three right-handed neutrinos which have zero charge under the
low energy gauged $U(1)_N$, and hence may get large masses.
In both versions additional singlet and vector-like states from $E_6$ reside
at the TeV scale, together with a $Z'$.
In the original version \cite{King:2005jy} extra vector-like Higgs states are added for the purposes
of unification, while in the minimal E$_6$SSM \cite{Howl:2007zi} they are not.

In this paper we will revisit an F-theory 
$E_6$ GUT model that has the TeV spectrum of the minimal E$_6$SSM,
namely three complete
27-dimensional representations of $E_6$ minus the right-handed neutrinos
\cite{Callaghan:2011jj,Callaghan:2012rv}
plus additional bulk exotics which provide the necessary states for unification
\cite{Callaghan:2013kaa}. Unification is achieved since the matter content 
is that of the MSSM supplemented
by four families of SU(5) $5+\bar{5}$ states, although in the present model
all the extra states are incomplete $SU(5)$ multiplets and crucially there are 
three additional TeV scale singlet states (in addition to the three high mass
right-handed neutrinos which are sufficient to realise the see-saw mechanism).
Moreover some of the low energy singlets couple
to three families of TeV scale vector-like matter with the quantum numbers of 
down-type quarks \cite{Callaghan:2012rv} called here $D,\bar{D}$. Unlike the E$_6$SSM,
the extra gauged  $U(1)_N$ may be broken at the GUT scale,
leading to an NMSSM-like theory without an extra $Z'$, but with
extra vector-like matter, as in the NMSSM+ \cite{Hall:2012mx}.
However, here we focus exclusively on the model in \cite{Callaghan:2013kaa}
where one of the three low energy singlets is responsible for the Higgs $\mu$ term,
and acquires an electroweak scale vacuum expectation value (VEV),
while the other two singlets do not couple to Higgs but do couple to vector-like 
quarks $D,\bar{D}$, acquiring a TeV scale VEV. These latter candidates are therefore candidates for 
the 750 GeV mass resonance, able to account for the ATLAS and CMS data,
since they have couplings to $D,\bar{D}$, and may have 
the required couplings required to generate 
the process $pp\to X\rightarrow \gamma\gamma$ via loops of $D,\bar{D}$ and inert Higgsinos.
We emphasise that these models were proposed before the recent ATLAS and CMS data,
so the interpretation that we discuss is not based on {\it ad hoc} modifications to the 
Standard Model, but rather represents a genuine consequence of well motivated
theoretical considerations.

The layout of the remainder of this paper is as follows. In the next section we review the
basic features of  the specific $E_6$ F-theory model 
focusing mainly on its spectrum and in particular on the properties of the predicted  
exotics. We start  section 3 by writing down the  Yukawa 
interactions related to the processes that interest us in this work. Next, we compute the
corresponding cross sections and compare our findings with the recent experimental results.
In section 4 we present our conclusions.

\section{The F-theory model with extra vector-like matter}

\begin{figure}[!ht]
\centering       
\includegraphics{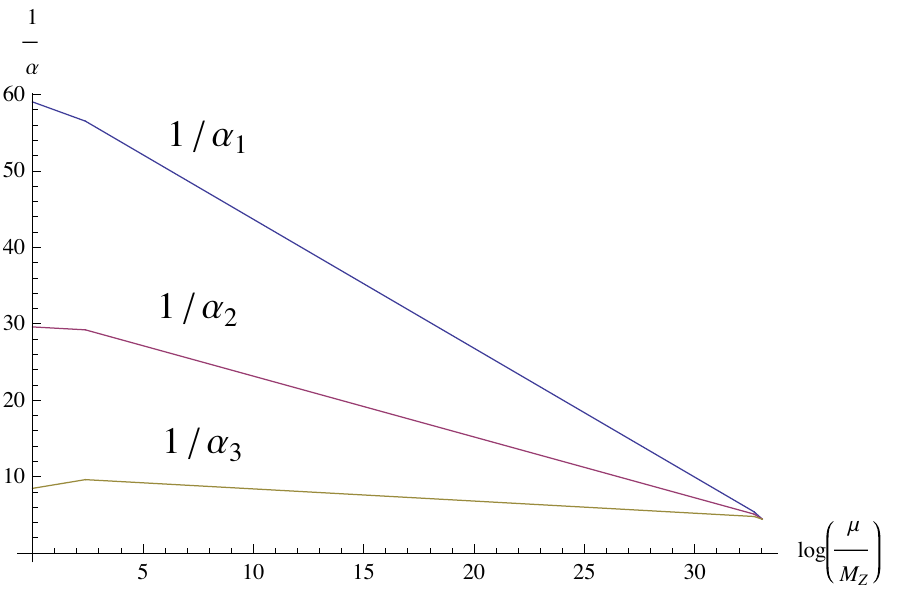}       
\caption{Gauge coupling unification in the model in Table~\ref{tevbulkspectrum} with TeV scale bulk exotics
with supersymmetry. The low energy matter content is that of 
the MSSM plus four extra $5+\overline{5}$ families of $SU(5)$,
although in fact these states originate from $E_6$ and consequently
in addition there are extra singlets which are responsible for the 750 GeV signal.}
\label{model2uni}
\end{figure}

In F-theory constructions SM-singlets and vector-like quark or lepton type fields are ubiquitous.
Many such pairs are expected to receive masses at a high scale, but it is possible that several of
them initially remain massless, later acquiring TeV scale masses.  
To set the stage, we start with a short description 
on the origin of the SM spectrum and bulk vector-like states in F-theory GUTs
in general. We choose $E_6$ as a working example
where it was shown sometime ago~\cite{Callaghan:2011jj,Callaghan:2012rv,Callaghan:2013kaa}
 that  scalars as well as vector-like fermion fields at the TeV scale are naturally accommodated. 
We start with the decomposition of the $E_8$-adjoint under the breaking  $E_8\supset E_6\times SU(3)$ 
 \[248\to (78,1)+(1,8)+(27,3)+(\ov{27},\bar 3)\]
 and label with $t_i$ the $SU(3)$ weights (subject to the tracelessness condition $t_1+t_2+t_3=0$). Along the $SU(3)$ 
 Cartan subalgebra, $(1,8)$ decomposes to singlets $\theta_{ij}, i,j=1,2,3$
  whilst the $27$'s 
 are characterised by the three charges $t_i$. We impose a  $Z_2$ monodromy $t_1=t_2$ thus, 
  we have the correspondence 
 \ba
 (1,8)\to  \theta_{13}, \theta_{31}, \theta_0;\;\ \ 
 (27,3)\to 27_{t_1},27_{t_3};\; \ \ (\overline{27},\bar 3)\to\overline{27}_{-t_1}, \overline{27}_{-t_3}
 \ea
Notice that because of the $Z_2$ monodromy we get the identifications $\theta_{12}= \theta_{21}\equiv \theta_0$,
as well as  $\theta_{23}=\theta_{13}$ and $\theta_{32}=\theta_{31}$
and analogously for the $27_{t_1}=27_{t_2}$. Additional bulk singlets $\theta_{kl}$ 
 and vector-like pairs are obtained under further breaking 
 of the symmetry down to $SU(5)$. 
The detailed derivation of the particular F-theory model we are interested in can be found
 in reference~\cite{Callaghan:2013kaa}. In the present note, we only present the $E_6$ origin 
 of the low energy spectrum   and the corresponding $SU(5)\times U(1)_N$ multiplets which are summarised in 
  Table~\ref{tevbulkspectrum}. The last column shows the `charge' $Q_N$ of the $U(1)_N$
  abelian gauge factor contained in $E_6$ under which the right-handed neutrinos are 
  singlets as in the E$_6$SSM \cite{King:2005jy}.
Without the bulk exotics the spectrum has the matter equivalent of three families of 
$E_6$ 27-dimensional representations as in the minimal E$_6$SSM \cite{Howl:2007zi},
which form an anomaly free set by themselves. Such a model was realised in F-theory context~\cite{Callaghan:2012rv} 
while it was shown that unification can be successfully achieved with the inclusion
of the bulk exotics~\cite{Callaghan:2013kaa} relevant to our present discussion. 
The total low energy spectrum, including bulk exotics,
then has the matter content of the MSSM plus four extra 
vector-like $5+\overline{5}$ families plus three extra 
singlets. Three right-handed neutrinos are present at high energies.
Renormalisation Group analysis shows~\cite{Callaghan:2013kaa} that perturbative unification can be achieved as shown in Fig.\ref{model2uni}.
With this in mind, next we focus on the characteristic properties of the model which are required to accommodate
the recent experimental data.

\begin{table}[ht]
\begin{center}
\small
\begin{tabular}{|c|c|c|c|}
\hline
$E_6$  & $SU(5)$  & TeV  spectrum & $\sqrt{10} Q_N$ \\ 
\hline
$27_{t_1}$ & $\ov{5}$ & $3(d^{c}+L)$ & $1$ \\
%\hline
$27_{t_1}$ & $10$ & $3(Q+u^{c}+e^{c})$ & $\frac{1}{2}$ \\
%\hline
$27_{t_1}$ & $5$ & $3D+2H_u$ & $-1$ \\
%\hline
$27_{t_1}$ & $\ov{5}$ & $3(\overline{D}+H_d)$ & $-\frac{3}{2}$ \\
%\hline
$27_{t_1}$ & $1$ & $\theta_{14}$ & $\frac{5}{2 }$ \\
%\hline
$27_{t_3}$ & $5$ & $H_u$ & $-\frac{1}{2 }$ \\
%\hline
$27_{t_3}$ & $1$ & $2\, \theta_{34}$ & $\frac{5}{2 }$ \\
%\hline
$78$ & $\ov{5}$ & $2\,X_{H_d} + X_{d^c}$  & $-\frac{3}{2 }$ \\
%\hline
$78$ & $5$ & $2\ov{X}_{\ov{H}_d} +  \ov{X}_{\ov{d^c}}$  & $\frac{3}{2}$ \\
1&1&$\theta_{13},\theta_{31},\theta_0$&0\\
\hline
\end{tabular}
\caption{\small The low energy spectrum for the F-theory
E$_6$SSM-like model with TeV scale bulk exotics
taken without change from~\cite{Callaghan:2013kaa}.
The fields $Q$, $u^c$, $d^c$, $L$, $e^c$ represent quark and lepton SM superfields in the usual notation.
In this spectrum there are three families of $H_u$ and $H_d$ Higgs superfields, as compared to a
 single one in the MSSM. There are also three families of exotic $D$ and $\overline{D}$ colour triplet superfields,
where $\overline{D}$ has the same SM quantum numbers as $d^c$, and $D$ has opposite quantum numbers.
We have written the bulk exotics as $X$ with a subscript that indicates the SM quantum numbers of that state.
The superfields $\theta_{ij}$ are SM singlets,
with the two $\theta_{34}$ singlets containing spin-0 candidates for 
the 750 GeV resonance. }
\label{tevbulkspectrum}
\end{center}
\end{table}%

\section{Production and decay of the 750 GeV scalar/pseudoscalar}

\begin{figure}[t!]\centering
\includegraphics[scale=0.4]{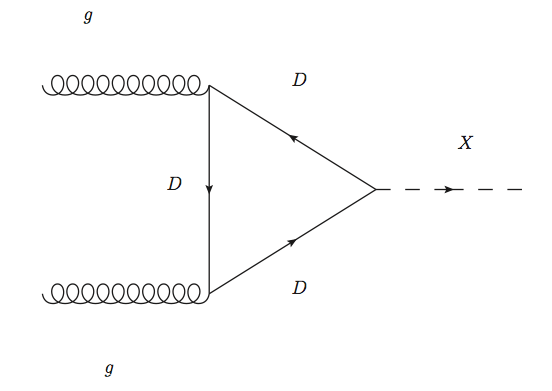}
\caption{The new singlet scalar/pseudoscalar $X\equiv \theta_{34}$ with mass 750 GeV is produced by gluon fusion due to its coupling to a loop
of vector-like fermions $D,\overline{D}$ which are colour triplets and have electric charge $\mp 1/3$.\label{gluon}}
\end{figure}

\begin{figure}[t!]\centering
\includegraphics[scale=0.4]{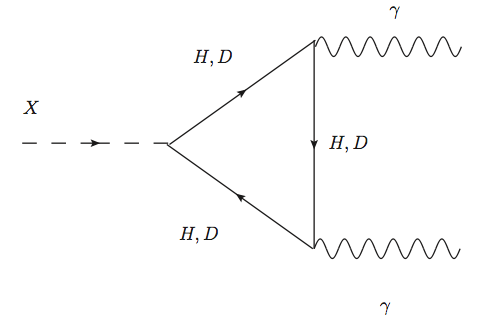}
\caption{The new singlet scalar/pseudoscalar $X\equiv \theta_{34}$ with mass 750 GeV is decays into a pair of photons due to its coupling to a loop
of vector-like fermions $H,\overline{H}$ which are colour singlet inert Higgsinos
with electric charge $\pm 1$ and 
$D,\overline{D}$ which are colour triplets and have electric charge $\mp 1/3$.\label{photon}}
\end{figure}

The terms in the superpotential which are responsible for generating the $\mu$ term and the exotic masses are~\cite{Callaghan:2013kaa}
\begin{equation}
W \sim \lambda \theta_{14} H_{d } H_{u } + \lambda_{\alpha \beta \gamma} \theta_{34}^{\alpha}
H_{d }^{\beta} H_{u }^{\gamma}
+ \kappa_{\alpha jk} \theta_{34}^{\alpha} \overline{D}_j D_k
\label{E6Y}
\end{equation}
These couplings originate from the $27_{t_1}27_{t_1}27_{t_3}$ E$_6$ coupling.
Thus two of the singlets $\theta_{34}^{\alpha}$  couple to all three of the colour triplet charge $\mp 1/3$
vector-like fermions $D_k,\ov{D}_j$ as well as two families of inert
Higgs doublets $H_{d }^{\beta}, H_{u }^{\gamma}$ (which do not get VEVs)
($\alpha ,  \beta , \gamma =1,2$).
One or both (if they are degenerate) singlet scalars may have a mass of 750 GeV and be produced by gluon
fusion at the LHC, decaying into two photons as shown in Figs.~\ref{gluon} and \ref{photon}.
A third singlet $\theta_{14}$ couples to the two Higgs doublets of the MSSM, and is responsible for
the effective $\mu$ term as in the NMSSM. However this singlet does not couple to coloured fermions
and so cannot be strongly produced at the LHC. It should also be mentioned in passing that the $E_6$ singlets can generate couplings such as $\theta_0 X_d\bar X_{\bar d}$ from the $E_6$ invariant term $78\cdot 78\cdot 1$, though we shall not discuss this further.

We therefore identify the 750 GeV scalar $S$ with a spin-0 component of one
of the F-theory singlets $\theta_{34}$,
which couples to three families of vector-like fermions $D_k,\ov{D}_j$
and two families of inert Higgs doublets $H_{d }^{\beta}, H_{u }^{\gamma}$.
The scalar components of $\theta_{34}$
are both assumed to develop TeV scale VEVs which are responsible
for generating the vector-like fermion masses for $D_k,\ov{D}_j$.
Since there are two complex singlets $\theta_{34}$, the spectrum will therefore
contain two scalars,
two pseudoscalars and two complex Weyl fermions. 
The two scalars plus two pseudoscalars are all candidates for the observed 
750 GeV resonance. If two or more of them are degenerate then this may lead to
an initially unresolved broad resonance. Eventually all four states may be discovered with 
different masses around the TeV scale, providing a smoking gun signature of the model.

\subsection{Cross Section}

We have seen that the spectrum of  the F-theory derived model contains complex singlet superfields  
possessing scalar and pseudoscalar components. 
The superpotential in Eq.\ref{E6Y},
below the scale of the VEVs of $X$ and the SUSY breaking scale,
gives rise to the low energy effective Lagrangian which contains terms like,
\[  {\cal L}\sim \kappa_iX \bar D_i D_i
+\lambda_{\alpha} X H_u^{\alpha}H_d^{\alpha}+M_i\bar D_i D_i
+M_{H_{\alpha}}H_u^{\alpha}H_d^{\alpha}
+\frac 12 M^2X^2 +\cdots \] %%%%%%%%
where $X$ is a scalar or pseudoscalar 
field originating from the $\theta_{34}$
coupled to a vector pair of fermions identified 
with the fermionic components of the three coloured triplet pairs $D_i,\bar D_i$,  
while $M_i$ are the three masses of 
the triplet fermions with $M_i\sim \kappa\langle \theta_{34} \rangle$ 
of~(\ref{E6Y}) and $M$ is the  mass of the singlet field
originating from a combination of soft SUSY breaking masses and 
the VEVs of the singlets. Similar couplings are also shown to the two families
of vector-like inert Higgsinos, labelled by $\alpha = 1,2$.

The vector-like fermions generate  loops diagrams which give rise to  Effective Field Theory 
d-5 operators. For the scalar component  $X\to S$ 
\be 
{\cal L}_{eff}\propto -\frac 14 S\left(g_{S\gamma} F_{\mu\nu}F^{\mu\nu}+g_{Sg} G_{\mu\nu}G^{\mu\nu}\right)
\ee
and analogously for pseudoscalar $X\to A$,
\be 
{\cal L}_{eff}\propto -\frac 14 A\left(g_{A\gamma} F_{\mu\nu}\tilde F^{\mu\nu}+g_{Ag} G_{\mu\nu}\tilde G^{\mu\nu}\right)\,.
\ee
A related mechanism has been already suggested as a plausible scenario in String derived
models~\cite{Cvetic:2015vit, Anchordoqui:2015jxc,Ibanez:2015uok}
where pseudoscalar fields such as axions and scalar fields such as the dilaton field have couplings of the above form. Here instead we regard the scalar and pseudoscalar as originating from a 27-dimensional
matter superfield, coupling to vector-like extra quarks which also appear in the 27-rep of $E_6$.

We consider a scalar/pseudoscalar particle $X$ originating from
one of the two $\theta_{34}$ fields, coupling to three families of colour triplet
charge $\mp 1/3$ extra vector-like quarks $D_i,\bar D_i$ and two families of Higgsinos 
$H_{u/d}^{\alpha}$ - as per Equation \ref{E6Y}.
The cross section for production of this scalar/pseudoscalar from gluon fusion, $\sigma(pp\rightarrow{X}\rightarrow\gamma\gamma)$, 
 where $X$ ia a uncoloured boson with mass M and spin $J=0$, can be written as \cite{Franceschini:2015kwy}

\begin{equation}\label{crsec}
	\sigma(pp\rightarrow{X}\rightarrow\gamma\gamma)=\frac{1}{M\Gamma s}C_{gg}\Gamma(X\rightarrow{gg})\Gamma(X\rightarrow{\gamma\gamma})
\end{equation}

\noindent where $C_{gg}$ is the dimensionless partonic integral for gluon production, which at 
$\sqrt{s}=13$ TeV is $C_{gg}=2137$. Here $\Gamma=\Gamma(X\rightarrow{gg})+\Gamma(X\rightarrow{\gamma\gamma})$ since no other interactions contribute to the effect.

 For the case in which a scalar/pseudoscalar resonance is produced from gluon fusion mediated by extra vector-like fermions $D_i,\bar D_i$ with mass $M_{i}$ and charges $Q_{i}$, decaying into two photons by a combination of the same vector-like fermions and Higgsinos $H_d^{\alpha}$ and $H_u^{\alpha}$, the corresponding decay widths read \cite{Franceschini:2015kwy}:

\begin{equation}\label{width1}
	\frac{\Gamma( X\rightarrow{gg})}{M}=\frac{\alpha_3^2}{2\pi^3}\left|\sum_{i}C_{r_{i}}\kappa_{i}\frac{2M_{i}}{M}\mathcal{X}\left(\frac{4M_i^2}{M^2}\right)\right|^{2},
\end{equation}

\begin{equation}\label{width2}
	\frac{\Gamma(X\rightarrow{\gamma\gamma})}{M}=\frac{\alpha^2}{16\pi^3}\left|\sum_{i}d_{r_{i}}Q^{2}_{i}\kappa_{i}\frac{2M_{i}}{M}\mathcal{X}\left(\frac{4M_i^2}{M^2}\right)+
	\sum_{\alpha}d_{r_{\alpha}}Q^{2}_{\alpha}
	\lambda_{\alpha}\frac{2M_{H_{\alpha}}}{M}\mathcal{X}\left(\frac{4M_{H{\alpha}}^2}{M^2}\right)\right|^{2}
\end{equation}

\begin{figure}[t!]
	\centering
	\includegraphics[scale=0.8]{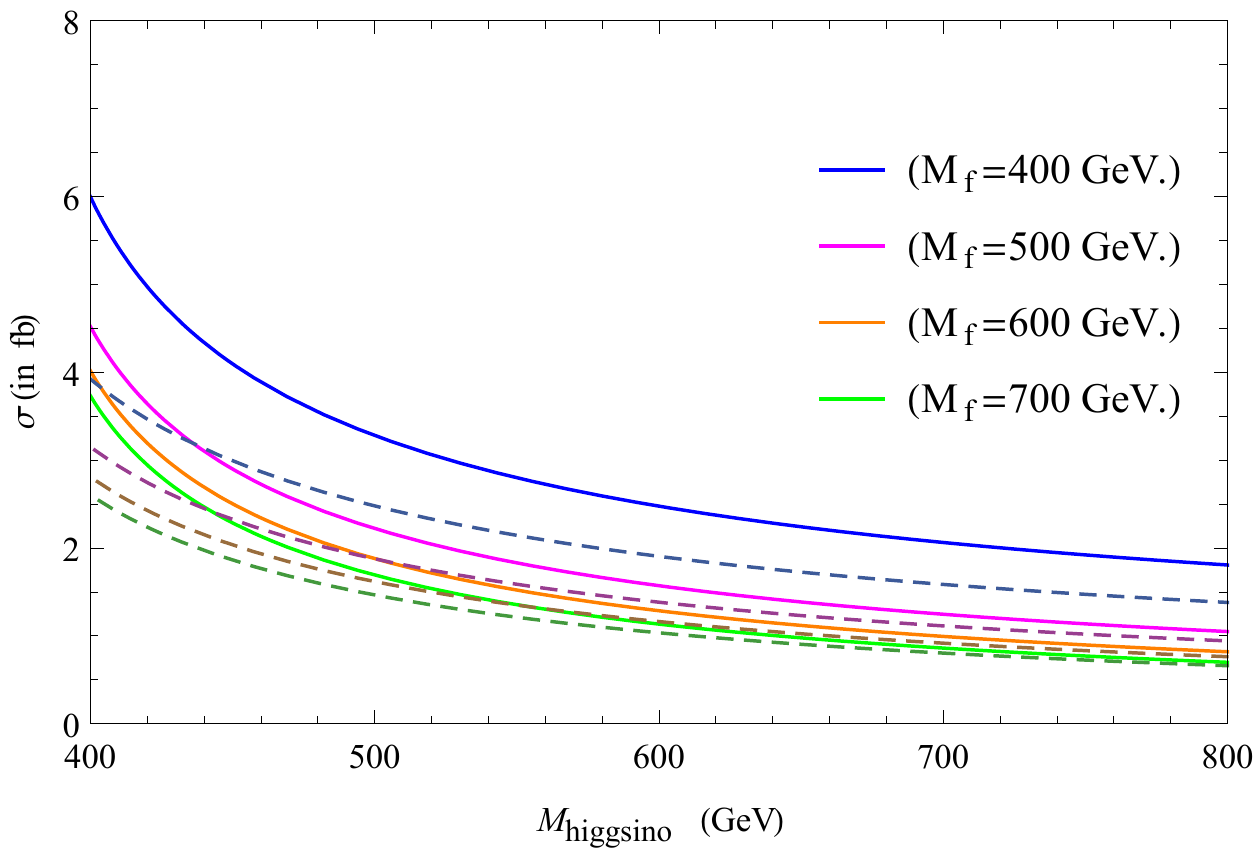}
	\caption{The cross section $\sigma(pp\rightarrow{X}\rightarrow\gamma\gamma)$ (in fb units) in the parametric space of the Higgsinos $H_u^{\beta}/H_d^{\gamma}$, for a selection of masses of the vector-like  $D_i/\overline{D}_i$ with all masses $M_i$ set equal to 
$M_{f}$ and the coupling $y_{f}$, with $y_{f}=1$. The solid lines correspond to the Pseudoscalar candidate state, while the dashed lines of the same hue correspond to the Scalar option.} \label{csplot} 
\end{figure}

\begin{figure}[t!]
	\centering
	\includegraphics[scale=0.8]{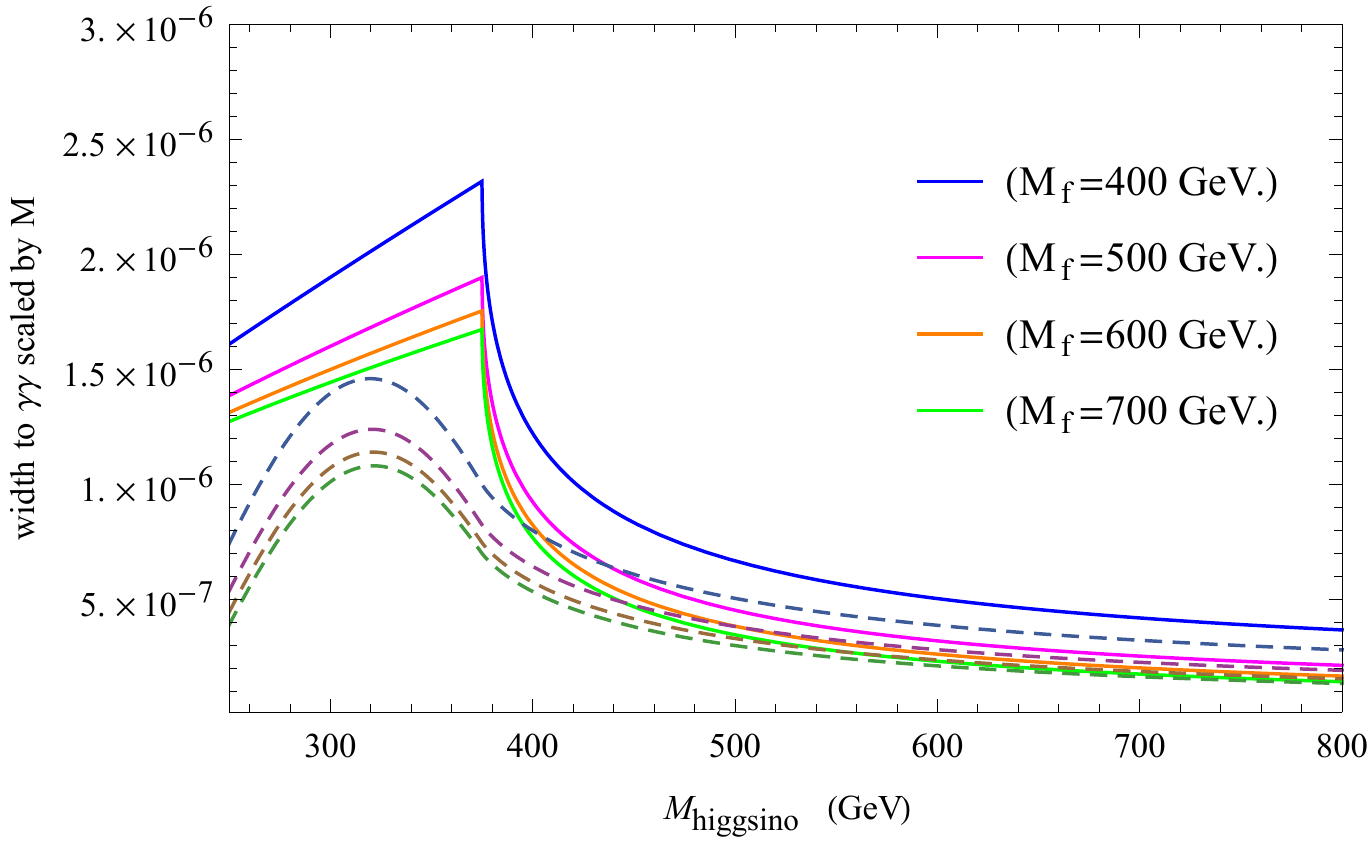}
	\caption{The mass weighted width $\Gamma(X\rightarrow\gamma\gamma)$ in the parametric space of the Higgsinos $H_u^{\beta}/H_d^{\gamma}$, for a selection of masses of the vector-like  $D_i/\overline{D}_i$ with masses $M_{f}$ and the coupling $y_{f}$, with $y_{f}=1$. The solid lines correspond to the Pseudoscalar candidate state, while the dashed lines of the same hue correspond to the Scalar option. } \label{widthplot} 
\end{figure}

The function $\mathcal{X}(t)$ takes a different form, depending on whether the particle is a scalar or a pseudoscalar - $\mathcal{S}$ or $\mathcal{P}$ respectively
\cite{Djouadi:2005gi}:
\begin{align}
\mathcal{P}(t)=\arctan^2(1/\sqrt{t-1}),\\
\mathcal{S}(t)=1+(1-t)\mathcal{P}(t)\,.
\end{align} 
In the case in question with colour triplets of mass $M_i$ mediating the process, $Q_{i}=1/3$, $C_{r_{i}}=1/2$, and $d_{r_{i}}=3$, while the Higgsinos have $Q_i=d_{r_i}=1$ and a mass of $M_k$. Combining the equations above we calculate the cross section for a scalar of mass $M=750$ GeV at $\sqrt{s}=13$ TeV. 
For simplicity we set all the masses of the vector-like fermions to be equal to 
degenerate (likewise for the Higgsinos), and all the couplings of the scalar singlet to the fermions to be equal to $y_f$. The results are presented in \fref{csplot} and \fref{widthplot}. Note also that the $\Gamma{(X\rightarrow{gg}})/M$ take values in the region of $10^{-4}$ and $10^{-5}$ which is not excluded by searches for dijet resonances at Run 1.

%The results are presented in \fref{contourplot} and \fref{fig2}, which is in the parameter space of the 
%remaining two free variables $y_f$ and $M_f$. Also for simplicity we consider the case 
%of just one scalar singlet $S$, whereas in fact the model predicts two scalars
%plus two pseudoscalars.

\section{Conclusions}
We have interpreted the 750-760 GeV diphoton resonance as one or more of
the spinless
components of two singlet superfields arising from the three
27-dimensional representations
of $E_6$ in F-theory,
which also contain three copies of colour-triplet charge $\mp 1/3$
vector-like
fermions $D_i,\bar{D}_i$ as well as inert
Higgsino doublets $H_{d }^{\beta}, H_{u }^{\gamma}$
to which the singlets may couple.
For definiteness we have considered (without change) a model that was
proposed some time ago
which contains such states, as well as bulk exotics, leading to gauge
coupling unification.

In order to obtain a large enough cross-section, we require the resonance
to be identified with one
of the two pseudoscalar (rather than scalar) states. However even in this
case, a sufficiently large cross-section
requires quite light colour triplets and charged Higgsinos below a TeV,
even with of order unit Yukawa couplings,
which is one of the predictions of the model.

The smoking gun prediction of the model is the existence of other similar
spinless
resonances, possibly close in mass to 750-760 GeV,
decaying into diphotons, as well as the three families of vector-like
fermions
$D_i,\bar{D}_i$ and two families of inert Higgsino doublets $H_{d
}^{\beta}, H_{u }^{\gamma}$.

It is possible that two or more of the singlet spinless states may be
close in mass, providing nearby resonances which
could be initially mistaken for a single broad resonance in the current
data. Indeed, from the 27 reps of the E$_6$ F-theory
model there are two singlet superfields which couple to the vector-like
fermions
$D_i,\bar{D}_i$, so there could be up to four spinless resonances which
can be searched for.

Further bulk singlets arising from the 78 reps of the E$_6$ F-theory
model are also expected to be present in the low energy spectrum whose
VEVs are responsible for the low energy exotic bulk
masses of the $2X_{H_d}$, $X_{d^c}$ and their vector partners. These bulk
singlets are also candidates for the 750 GeV diphoton
resonance, or may have similar masses.

In conclusion, realistic E$_6$ F-theory models generically contain extra
low energy states which include a plethora of spinless singlets
and vector-like fermions with various charges and colours, especially
colour singlet unit charged states and colour triplets
with charges $\mp 1/3$, which appear to have the correct properties to
provide an explanation of the 750 GeV diphoton
resonance indicated by the LHC Run 2 data. We have discussed an already
existing model (without change) which is perfectly
capable of accounting for these data, as well as furnishing many
predictions of multiple other similar resonances as
well as the exotic fermions and their superpartners which should be
observable in future.

\vfill

\vspace{0.1in}
SFK acknowledges partial support from the STFC Consolidated ST/J000396/1 grant and 
the European Union FP7 ITN-INVISIBLES (Marie Curie Actions, PITN-
GA-2011-289442). AKM is supported by an STFC studentship.


\vspace{0.1in}

\newpage


\begin{thebibliography}{99}

\bibitem{ATLAS}
ATLAS Collaboration, ``Search for resonances decaying to photon pairs in 3.2 fb$^{?1}$ of pp collisions at $\sqrt{s}$ = 13 TeV with the ATLAS detector'', ATLAS-CONF-2015- 081.

\bibitem{CMS}
CMS Collaboration, ``Search for new physics in high mass diphoton events in proton-proton collisions at $\sqrt{s}$  = 13 TeV'', CMS-PAS-EXO-15-004.

  
  
 %\cite{Harigaya:2015ezk}
 \bibitem{Harigaya:2015ezk}
   K.~Harigaya and Y.~Nomura,
   %``Composite Models for the 750 GeV Diphoton Excess,''
   arXiv:1512.04850 [hep-ph].
   %%CITATION = ARXIV:1512.04850;%%
   %80 citations counted in INSPIRE as of 30 Dec 2015
   
     
   %\cite{Mambrini:2015wyu}
\bibitem{Mambrini:2015wyu}
  Y.~Mambrini, G.~Arcadi and A.~Djouadi,
  %``The LHC diphoton resonance and dark matter,''
  arXiv:1512.04913 [hep-ph].
  %%CITATION = ARXIV:1512.04913;%%
  %104 citations counted in INSPIRE as of 07 Jan 2016
  
  %\cite{Backovic:2015fnp}
  \bibitem{Backovic:2015fnp}
  M.~Backovic, A.~Mariotti and D.~Redigolo,
  %``Di-photon excess illuminates Dark Matter,''
  arXiv:1512.04917 [hep-ph].
  %%CITATION = ARXIV:1512.04917;%%
  %85 citations counted in INSPIRE as of 04 Jan 2016

  
  %\cite{Angelescu:2015uiz}
\bibitem{Angelescu:2015uiz}
  A.~Angelescu, A.~Djouadi and G.~Moreau,
  %``Scenarii for interpretations of the LHC diphoton excess: two Higgs doublets and vector-like quarks and leptons,''
  arXiv:1512.04921 [hep-ph].
  %%CITATION = ARXIV:1512.04921;%%
  %103 citations counted in INSPIRE as of 07 Jan 2016
  
  
   %\cite{Nakai:2015ptz}
\bibitem{Nakai:2015ptz}
  Y.~Nakai, R.~Sato and K.~Tobioka,
  %``Footprints of New Strong Dynamics via Anomaly,''
  arXiv:1512.04924 [hep-ph].
  %%CITATION = ARXIV:1512.04924;%%
  %86 citations counted in INSPIRE as of 07 Jan 2016

   
   %\cite{Knapen:2015dap}
\bibitem{Knapen:2015dap}
  S.~Knapen, T.~Melia, M.~Papucci and K.~Zurek,
  %``Rays of light from the LHC,''
  arXiv:1512.04928 [hep-ph].
  %%CITATION = ARXIV:1512.04928;%%
  %99 citations counted in INSPIRE as of 07 Jan 2016
   
   %\cite{Pilaftsis:2015ycr}
   \bibitem{Pilaftsis:2015ycr}
     A.~Pilaftsis,
 %    ``Diphoton Signatures from Heavy Axion Decays at LHC,''
     arXiv:1512.04931 [hep-ph].
     %%CITATION = ARXIV:1512.04931;%%
     %66 citations counted in INSPIRE as of 25 Dec 2015
   
   
   
 %\cite{Franceschini:2015kwy}
\bibitem{Franceschini:2015kwy} 
 R.~Franceschini {\it et al.},
 %``What is the gamma gamma resonance at 750 GeV?,''
  arXiv:1512.04933 [hep-ph].
 %%CITATION = ARXIV:1512.04933;%%
  %94 citations counted in INSPIRE as of 30 Dec 2015
  
  %\cite{DiChiara:2015vdm}
\bibitem{DiChiara:2015vdm}
  S.~Di Chiara, L.~Marzola and M.~Raidal,
  %``First interpretation of the 750 GeV di-photon resonance at the LHC,''
  arXiv:1512.04939 [hep-ph].
  %%CITATION = ARXIV:1512.04939;%%
  %102 citations counted in INSPIRE as of 07 Jan 2016
  
    %\cite{Ellis:2015oso}
  \bibitem{Ellis:2015oso}
  J.~Ellis, S.~A.~R.~Ellis, J.~Quevillon, V.~Sanz and T.~You,
  %``On the Interpretation of a Possible $\sim 750$ GeV Particle Decaying into $\gamma \gamma$,''
  arXiv:1512.05327 [hep-ph].
  %%CITATION = ARXIV:1512.05327;%%
  %86 citations counted in INSPIRE as of 04 Jan 2016
  
  %\cite{Bellazzini:2015nxw}
\bibitem{Bellazzini:2015nxw}
  B.~Bellazzini, R.~Franceschini, F.~Sala and J.~Serra,
  %``Goldstones in Diphotons,''
  arXiv:1512.05330 [hep-ph].
  %%CITATION = ARXIV:1512.05330;%%
  %89 citations counted in INSPIRE as of 07 Jan 2016
  
  %\cite{Gupta:2015zzs}
  \bibitem{Gupta:2015zzs}
  R.~S.~Gupta, S.~Jäger, Y.~Kats, G.~Perez and E.~Stamou,
  %``Interpreting a 750 GeV Diphoton Resonance,''
  arXiv:1512.05332 [hep-ph].
  %%CITATION = ARXIV:1512.05332;%%
  %82 citations counted in INSPIRE as of 04 Jan 2016
  %\cite{Higaki:2015jag}
  \bibitem{Higaki:2015jag}
  T.~Higaki, K.~S.~Jeong, N.~Kitajima and F.~Takahashi,
  %``The QCD Axion from Aligned Axions and Diphoton Excess,''
  arXiv:1512.05295 [hep-ph].
  %%CITATION = ARXIV:1512.05295;%%
  %73 citations counted in INSPIRE as of 04 Jan 2016
  %\cite{McDermott:2015sck}
  \bibitem{McDermott:2015sck}
  S.~D.~McDermott, P.~Meade and H.~Ramani,
  %``Singlet Scalar Resonances and the Diphoton Excess,''
  arXiv:1512.05326 [hep-ph].
  %%CITATION = ARXIV:1512.05326;%%
  %87 citations counted in INSPIRE as of 04 Jan 2016
  %\cite{Low:2015qep}
  \bibitem{Low:2015qep}
  M.~Low, A.~Tesi and L.~T.~Wang,
  %``A pseudoscalar decaying to photon pairs in the early LHC run 2 data,''
  arXiv:1512.05328 [hep-ph].
  %%CITATION = ARXIV:1512.05328;%%
  %86 citations counted in INSPIRE as of 04 Jan 2016
  %\cite{Petersson:2015mkr}
  \bibitem{Petersson:2015mkr}
  C.~Petersson and R.~Torre,
  %``The 750 GeV diphoton excess from the goldstino superpartner,''
  arXiv:1512.05333 [hep-ph].
  %%CITATION = ARXIV:1512.05333;%%
  %82 citations counted in INSPIRE as of 04 Jan 2016
  
  %\cite{Molinaro:2015cwg}
\bibitem{Molinaro:2015cwg}
  E.~Molinaro, F.~Sannino and N.~Vignaroli,
  %``Minimal Composite Dynamics versus Axion Origin of the Diphoton excess,''
  arXiv:1512.05334 [hep-ph].
  %%CITATION = ARXIV:1512.05334;%%
  %88 citations counted in INSPIRE as of 07 Jan 2016
  
  
  %\cite{Dutta:2015wqh}
  \bibitem{Dutta:2015wqh}
  B.~Dutta, Y.~Gao, T.~Ghosh, I.~Gogoladze and T.~Li,
  %``Interpretation of the diphoton excess at CMS and ATLAS,''
  arXiv:1512.05439 [hep-ph].
  %%CITATION = ARXIV:1512.05439;%%
  %66 citations counted in INSPIRE as of 04 Jan 2016
  %\cite{Cao:2015pto}
  \bibitem{Cao:2015pto}
  Q.~H.~Cao, Y.~Liu, K.~P.~Xie, B.~Yan and D.~M.~Zhang,
  %``A Boost Test of Anomalous Diphoton Resonance at the LHC,''
  arXiv:1512.05542 [hep-ph].
  %%CITATION = ARXIV:1512.05542;%%
  %62 citations counted in INSPIRE as of 04 Jan 2016
  %\cite{Kobakhidze:2015ldh}
  \bibitem{Kobakhidze:2015ldh}
  A.~Kobakhidze, F.~Wang, L.~Wu, J.~M.~Yang and M.~Zhang,
  %``LHC 750 GeV diphoton resonance explained as a heavy scalar in top-seesaw model,''
  arXiv:1512.05585 [hep-ph].
  %%CITATION = ARXIV:1512.05585;%%
  %68 citations counted in INSPIRE as of 04 Jan 2016
  %\cite{Cox:2015ckc}
  \bibitem{Cox:2015ckc}
  P.~Cox, A.~D.~Medina, T.~S.~Ray and A.~Spray,
  %``Diphoton Excess at 750 GeV from a Radion in the Bulk-Higgs Scenario,''
  arXiv:1512.05618 [hep-ph].
  %%CITATION = ARXIV:1512.05618;%%
  %67 citations counted in INSPIRE as of 04 Jan 2016
  %\cite{Ahmed:2015uqt}
  \bibitem{Ahmed:2015uqt}
  A.~Ahmed, B.~M.~Dillon, B.~Grzadkowski, J.~F.~Gunion and Y.~Jiang,
  %``Higgs-radion interpretation of 750 GeV di-photon excess at the LHC,''
  arXiv:1512.05771 [hep-ph].
  %%CITATION = ARXIV:1512.05771;%%
  %62 citations counted in INSPIRE as of 04 Jan 2016
  %\cite{Becirevic:2015fmu}
  \bibitem{Becirevic:2015fmu}
  D.~Becirevic, E.~Bertuzzo, O.~Sumensari and R.~Z.~Funchal,
  %``Can the new resonance at LHC be a CP-Odd Higgs boson?,''
  arXiv:1512.05623 [hep-ph].
  %%CITATION = ARXIV:1512.05623;%%
  %62 citations counted in INSPIRE as of 04 Jan 2016
  %\cite{No:2015bsn}
  \bibitem{No:2015bsn}
  J.~M.~No, V.~Sanz and J.~Setford,
  %``See-Saw Composite Higgses at the LHC: Linking Naturalness to the $750$ GeV Di-Photon Resonance,''
  arXiv:1512.05700 [hep-ph].
  %%CITATION = ARXIV:1512.05700;%%
  %68 citations counted in INSPIRE as of 04 Jan 2016
  %\cite{Demidov:2015zqn}
  \bibitem{Demidov:2015zqn}
  S.~V.~Demidov and D.~S.~Gorbunov,
  %``On sgoldstino interpretation of the diphoton excess,''
  arXiv:1512.05723 [hep-ph].
  %%CITATION = ARXIV:1512.05723;%%
  %63 citations counted in INSPIRE as of 04 Jan 2016
  %\cite{Chao:2015ttq}
  \bibitem{Chao:2015ttq}
  W.~Chao, R.~Huo and J.~H.~Yu,
  %``The Minimal Scalar-Stealth Top Interpretation of the Diphoton Excess,''
  arXiv:1512.05738 [hep-ph].
  %%CITATION = ARXIV:1512.05738;%%
  %68 citations counted in INSPIRE as of 04 Jan 2016
  %\cite{Fichet:2015vvy}
  \bibitem{Fichet:2015vvy}
  S.~Fichet, G.~von Gersdorff and C.~Royon,
  %``Scattering Light by Light at 750 GeV at the LHC,''
  arXiv:1512.05751 [hep-ph].
  %%CITATION = ARXIV:1512.05751;%%
  %67 citations counted in INSPIRE as of 04 Jan 2016
  %\cite{Curtin:2015jcv}
  \bibitem{Curtin:2015jcv}
  D.~Curtin and C.~B.~Verhaaren,
  %``Quirky Explanations for the Diphoton Excess,''
  arXiv:1512.05753 [hep-ph].
  %%CITATION = ARXIV:1512.05753;%%
  %62 citations counted in INSPIRE as of 04 Jan 2016
  %\cite{Bian:2015kjt}
  \bibitem{Bian:2015kjt}
  L.~Bian, N.~Chen, D.~Liu and J.~Shu,
  %``A hidden confining world on the 750 GeV diphoton excess,''
  arXiv:1512.05759 [hep-ph].
  %%CITATION = ARXIV:1512.05759;%%
  %69 citations counted in INSPIRE as of 04 Jan 2016
  %\cite{Chakrabortty:2015hff}
  \bibitem{Chakrabortty:2015hff}
  J.~Chakrabortty, A.~Choudhury, P.~Ghosh, S.~Mondal and T.~Srivastava,
  %``Di-photon resonance around 750 GeV: shedding light on the theory underneath,''
  arXiv:1512.05767 [hep-ph].
  %%CITATION = ARXIV:1512.05767;%%
  %65 citations counted in INSPIRE as of 04 Jan 2016
  %\cite{Csaki:2015vek}
  \bibitem{Csaki:2015vek}
  C.~Csaki, J.~Hubisz and J.~Terning,
  %``The Minimal Model of a Diphoton Resonance: Production without Gluon Couplings,''
  arXiv:1512.05776 [hep-ph].
  %%CITATION = ARXIV:1512.05776;%%
  %66 citations counted in INSPIRE as of 04 Jan 2016
  %\cite{Falkowski:2015swt}
  \bibitem{Falkowski:2015swt}
  A.~Falkowski, O.~Slone and T.~Volansky,
  %``Phenomenology of a 750 GeV Singlet,''
  arXiv:1512.05777 [hep-ph].
  %%CITATION = ARXIV:1512.05777;%%
  %73 citations counted in INSPIRE as of 04 Jan 2016
  %\cite{Bai:2015nbs}
  \bibitem{Bai:2015nbs}
  Y.~Bai, J.~Berger and R.~Lu,
  %``A 750 GeV Dark Pion: Cousin of a Dark G-parity-odd WIMP,''
  arXiv:1512.05779 [hep-ph].
  %%CITATION = ARXIV:1512.05779;%%
  %63 citations counted in INSPIRE as of 04 Jan 2016
  %\cite{Benbrik:2015fyz}
  \bibitem{Benbrik:2015fyz}
  R.~Benbrik, C.~H.~Chen and T.~Nomura,
  %``Higgs singlet as a diphoton resonance in a vector-like quark model,''
  arXiv:1512.06028 [hep-ph].
  %%CITATION = ARXIV:1512.06028;%%
  %45 citations counted in INSPIRE as of 04 Jan 2016
  %\cite{Kim:2015ron}
  \bibitem{Kim:2015ron}
  J.~S.~Kim, J.~Reuter, K.~Rolbiecki and R.~R.~de Austri,
  %``A resonance without resonance: scrutinizing the diphoton excess at 750 GeV,''
  arXiv:1512.06083 [hep-ph].
  %%CITATION = ARXIV:1512.06083;%%
  %39 citations counted in INSPIRE as of 04 Jan 2016
  
  %\cite{Gabrielli:2015dhk}
  \bibitem{Gabrielli:2015dhk}
  E.~Gabrielli, K.~Kannike, B.~Mele, M.~Raidal, C.~Spethmann and H.~Veermäe,
  %``A SUSY Inspired Simplified Model for the 750 GeV Diphoton Excess,''
  arXiv:1512.05961 [hep-ph].
  %%CITATION = ARXIV:1512.05961;%%
  %44 citations counted in INSPIRE as of 04 Jan 2016
  %\cite{Alves:2015jgx}
  \bibitem{Alves:2015jgx}
  A.~Alves, A.~G.~Dias and K.~Sinha,
  %``The 750 GeV $S$-cion: Where else should we look for it?,''
  arXiv:1512.06091 [hep-ph].
  %%CITATION = ARXIV:1512.06091;%%
  %48 citations counted in INSPIRE as of 04 Jan 2016
  %\cite{Megias:2015ory}
  \bibitem{Megias:2015ory}
  E.~Megias, O.~Pujolas and M.~Quiros,
  %``On dilatons and the LHC diphoton excess,''
  arXiv:1512.06106 [hep-ph].
  %%CITATION = ARXIV:1512.06106;%%
  %43 citations counted in INSPIRE as of 04 Jan 2016
  %\cite{Carpenter:2015ucu}
  \bibitem{Carpenter:2015ucu}
  L.~M.~Carpenter, R.~Colburn and J.~Goodman,
  %``Supersoft SUSY Models and the 750 GeV Diphoton Excess, Beyond Effective Operators,''
  arXiv:1512.06107 [hep-ph].
  %%CITATION = ARXIV:1512.06107;%%
  %42 citations counted in INSPIRE as of 04 Jan 2016
  %\cite{Bernon:2015abk}
  \bibitem{Bernon:2015abk}
  J.~Bernon and C.~Smith,
  %``Could the width of the diphoton anomaly signal a three-body decay ?,''
  arXiv:1512.06113 [hep-ph].
  %%CITATION = ARXIV:1512.06113;%%
  %39 citations counted in INSPIRE as of 04 Jan 2016
  %\cite{Chao:2015nsm}
  \bibitem{Chao:2015nsm}
  W.~Chao,
  %``Symmetries Behind the 750 GeV Diphoton Excess,''
  arXiv:1512.06297 [hep-ph].
  %%CITATION = ARXIV:1512.06297;%%
  %36 citations counted in INSPIRE as of 04 Jan 2016
  %\cite{Han:2015cty}
  \bibitem{Han:2015cty}
  C.~Han, H.~M.~Lee, M.~Park and V.~Sanz,
  %``The diphoton resonance as a gravity mediator of dark matter,''
  arXiv:1512.06376 [hep-ph].
  %%CITATION = ARXIV:1512.06376;%%
  %34 citations counted in INSPIRE as of 04 Jan 2016
  %\cite{Chang:2015bzc}
  \bibitem{Chang:2015bzc}
  S.~Chang,
  %``A Simple $U(1)$ Gauge Theory Explanation of the Diphoton Excess,''
  arXiv:1512.06426 [hep-ph].
  %%CITATION = ARXIV:1512.06426;%%
  %29 citations counted in INSPIRE as of 04 Jan 2016
  %\cite{Dhuria:2015ufo}
  \bibitem{Dhuria:2015ufo}
  M.~Dhuria and G.~Goswami,
  %``Perturbativity, vacuum stability and inflation in the light of 750 GeV diphoton excess,''
  arXiv:1512.06782 [hep-ph].
  %%CITATION = ARXIV:1512.06782;%%
  %27 citations counted in INSPIRE as of 04 Jan 2016
  %\cite{Han:2015dlp}
  \bibitem{Han:2015dlp}
  H.~Han, S.~Wang and S.~Zheng,
  %``Scalar Explanation of Diphoton Excess at LHC,''
  arXiv:1512.06562 [hep-ph].
  %%CITATION = ARXIV:1512.06562;%%
  %28 citations counted in INSPIRE as of 04 Jan 2016
  %\cite{Luo:2015yio}
  \bibitem{Luo:2015yio}
  M.~x.~Luo, K.~Wang, T.~Xu, L.~Zhang and G.~Zhu,
  %``Squarkonium/Diquarkonium and the Di-photon Excess,''
  arXiv:1512.06670 [hep-ph].
  %%CITATION = ARXIV:1512.06670;%%
  %25 citations counted in INSPIRE as of 04 Jan 2016
  %\cite{Chang:2015sdy}
  \bibitem{Chang:2015sdy}
  J.~Chang, K.~Cheung and C.~T.~Lu,
  %``Interpreting the 750 GeV Di-photon Resonance using photon-jets in Hidden-Valley-like models,''
  arXiv:1512.06671 [hep-ph].
  %%CITATION = ARXIV:1512.06671;%%
  %30 citations counted in INSPIRE as of 04 Jan 2016
  %\cite{Bardhan:2015hcr}
  \bibitem{Bardhan:2015hcr}
  D.~Bardhan, D.~Bhatia, A.~Chakraborty, U.~Maitra, S.~Raychaudhuri and T.~Samui,
  %``Radion Candidate for the LHC Diphoton Resonance,''
  arXiv:1512.06674 [hep-ph].
  %%CITATION = ARXIV:1512.06674;%%
  %31 citations counted in INSPIRE as of 04 Jan 2016
  %\cite{Feng:2015wil}
  \bibitem{Feng:2015wil}
  T.~F.~Feng, X.~Q.~Li, H.~B.~Zhang and S.~M.~Zhao,
  %``The LHC 750 GeV diphoton excess in supersymmetry with gauged baryon and lepton numbers,''
  arXiv:1512.06696 [hep-ph].
  %%CITATION = ARXIV:1512.06696;%%
  %33 citations counted in INSPIRE as of 04 Jan 2016
  %\cite{Liao:2015tow}
  \bibitem{Liao:2015tow}
  W.~Liao and H.~q.~Zheng,
  %``Scalar resonance at 750 GeV as composite of heavy vector-like fermions,''
  arXiv:1512.06741 [hep-ph].
  %%CITATION = ARXIV:1512.06741;%%
  %28 citations counted in INSPIRE as of 04 Jan 2016
  %\cite{Cho:2015nxy}
  \bibitem{Cho:2015nxy}
  W.~S.~Cho, D.~Kim, K.~Kong, S.~H.~Lim, K.~T.~Matchev, J.~C.~Park and M.~Park,
  %``The 750 GeV Diphoton Excess May Not Imply a 750 GeV Resonance,''
  arXiv:1512.06824 [hep-ph].
  %%CITATION = ARXIV:1512.06824;%%
  %34 citations counted in INSPIRE as of 04 Jan 2016
  %\cite{Barducci:2015gtd}
  \bibitem{Barducci:2015gtd}
  D.~Barducci, A.~Goudelis, S.~Kulkarni and D.~Sengupta,
  %``One jet to rule them all: monojet constraints and invisible decays of a 750 GeV diphoton resonance,''
  arXiv:1512.06842 [hep-ph].
  %%CITATION = ARXIV:1512.06842;%%
  %34 citations counted in INSPIRE as of 04 Jan 2016
  %\cite{Ding:2015rxx}
  \bibitem{Ding:2015rxx}
  R.~Ding, L.~Huang, T.~Li and B.~Zhu,
  %``Interpreting $750$ GeV Diphoton Excess with R-parity Violation Supersymmetry,''
  arXiv:1512.06560 [hep-ph].
  %%CITATION = ARXIV:1512.06560;%%
  %35 citations counted in INSPIRE as of 04 Jan 2016
  %\cite{Han:2015qqj}
  \bibitem{Han:2015qqj}
  X.~F.~Han and L.~Wang,
  %``Implication of the 750 GeV diphoton resonance on two-Higgs-doublet model and its extensions with Higgs field,''
  arXiv:1512.06587 [hep-ph].
  %%CITATION = ARXIV:1512.06587;%%
  %34 citations counted in INSPIRE as of 04 Jan 2016
  %\cite{Antipin:2015kgh}
  \bibitem{Antipin:2015kgh}
  O.~Antipin, M.~Mojaza and F.~Sannino,
  %``A natural Coleman-Weinberg theory explains the diphoton excess,''
  arXiv:1512.06708 [hep-ph].
  %%CITATION = ARXIV:1512.06708;%%
  %28 citations counted in INSPIRE as of 04 Jan 2016
  %\cite{Wang:2015kuj}
  \bibitem{Wang:2015kuj}
  F.~Wang, L.~Wu, J.~M.~Yang and M.~Zhang,
  %``750 GeV Diphoton Resonance, 125 GeV Higgs and Muon g-2 Anomaly in Deflected Anomaly Mediation SUSY Breaking Scenario,''
  arXiv:1512.06715 [hep-ph].
  %%CITATION = ARXIV:1512.06715;%%
  %35 citations counted in INSPIRE as of 04 Jan 2016
  %\cite{Cao:2015twy}
  \bibitem{Cao:2015twy}
  J.~Cao, C.~Han, L.~Shang, W.~Su, J.~M.~Yang and Y.~Zhang,
  %``Interpreting the 750 GeV diphoton excess by the singlet extension of the Manohar-Wise Model,''
  arXiv:1512.06728 [hep-ph].
  %%CITATION = ARXIV:1512.06728;%%
  %38 citations counted in INSPIRE as of 04 Jan 2016
  %\cite{Huang:2015evq}
  \bibitem{Huang:2015evq}
  F.~P.~Huang, C.~S.~Li, Z.~L.~Liu and Y.~Wang,
  %``750 GeV Diphoton Excess from Cascade Decay,''
  arXiv:1512.06732 [hep-ph].
  %%CITATION = ARXIV:1512.06732;%%
  %36 citations counted in INSPIRE as of 04 Jan 2016
  %\cite{Heckman:2015kqk}
  \bibitem{Heckman:2015kqk}
  J.~J.~Heckman,
  %``750 GeV Diphotons from a D3-brane,''
  arXiv:1512.06773 [hep-ph].
  %%CITATION = ARXIV:1512.06773;%%
  %31 citations counted in INSPIRE as of 04 Jan 2016
  %\cite{Bi:2015uqd}
  \bibitem{Bi:2015uqd}
  X.~J.~Bi, Q.~F.~Xiang, P.~F.~Yin and Z.~H.~Yu,
  %``The 750 GeV diphoton excess at the LHC and dark matter constraints,''
  arXiv:1512.06787 [hep-ph].
  %%CITATION = ARXIV:1512.06787;%%
  %35 citations counted in INSPIRE as of 04 Jan 2016
  
  %\cite{Kim:2015ksf}
  \bibitem{Kim:2015ksf}
  J.~S.~Kim, K.~Rolbiecki and R.~R.~de Austri,
  %``Model-independent combination of diphoton constraints at 750 GeV,''
  arXiv:1512.06797 [hep-ph].
  %%CITATION = ARXIV:1512.06797;%%
  %31 citations counted in INSPIRE as of 04 Jan 2016
  %\cite{Berthier:2015vbb}
  \bibitem{Berthier:2015vbb}
  L.~Berthier, J.~M.~Cline, W.~Shepherd and M.~Trott,
  %``Effective interpretations of a diphoton excess,''
  arXiv:1512.06799 [hep-ph].
  %%CITATION = ARXIV:1512.06799;%%
  %31 citations counted in INSPIRE as of 04 Jan 2016
  %\cite{Cline:2015msi}
  \bibitem{Cline:2015msi}
  J.~M.~Cline and Z.~Liu,
  %``LHC diphotons from electroweakly pair-produced composite pseudoscalars,''
  arXiv:1512.06827 [hep-ph].
  %%CITATION = ARXIV:1512.06827;%%
  %29 citations counted in INSPIRE as of 04 Jan 2016
  %\cite{Bauer:2015boy}
  \bibitem{Bauer:2015boy}
  M.~Bauer and M.~Neubert,
  %``Flavor Anomalies, the Diphoton Excess and a Dark Matter Candidate,''
  arXiv:1512.06828 [hep-ph].
  %%CITATION = ARXIV:1512.06828;%%
  %33 citations counted in INSPIRE as of 04 Jan 2016
  %\cite{Chala:2015cev}
  \bibitem{Chala:2015cev}
  M.~Chala, M.~Duerr, F.~Kahlhoefer and K.~Schmidt-Hoberg,
  %``Tricking Landau-Yang: How to obtain the diphoton excess from a vector resonance,''
  arXiv:1512.06833 [hep-ph].
  %%CITATION = ARXIV:1512.06833;%%
  %29 citations counted in INSPIRE as of 04 Jan 2016
  %\cite{Boucenna:2015pav}
  \bibitem{Boucenna:2015pav}
  S.~M.~Boucenna, S.~Morisi and A.~Vicente,
  %``The LHC diphoton resonance from gauge symmetry,''
  arXiv:1512.06878 [hep-ph].
  %%CITATION = ARXIV:1512.06878;%%
  %26 citations counted in INSPIRE as of 04 Jan 2016
  %\cite{Dev:2015isx}
  \bibitem{Dev:2015isx}
  P.~S.~B.~Dev and D.~Teresi,
  %``Asymmetric Dark Matter in the Sun and the Diphoton Excess at the LHC,''
  arXiv:1512.07243 [hep-ph].
  %%CITATION = ARXIV:1512.07243;%%
  %29 citations counted in INSPIRE as of 04 Jan 2016
  %\cite{deBlas:2015hlv}
  \bibitem{deBlas:2015hlv}
  J.~de Blas, J.~Santiago and R.~Vega-Morales,
  %``New vector bosons and the diphoton excess,''
  arXiv:1512.07229 [hep-ph].
  %%CITATION = ARXIV:1512.07229;%%
  %26 citations counted in INSPIRE as of 04 Jan 2016
  %\cite{Murphy:2015kag}
  \bibitem{Murphy:2015kag}
  C.~W.~Murphy,
  %``Vector Leptoquarks and the 750 GeV Diphoton Resonance at the LHC,''
  arXiv:1512.06976 [hep-ph].
  %%CITATION = ARXIV:1512.06976;%%
  %30 citations counted in INSPIRE as of 04 Jan 2016
  %\cite{Hernandez:2015ywg}
  \bibitem{Hernandez:2015ywg}
  A.~E.~C.~Hernández and I.~Nisandzic,
  %``LHC diphoton 750 GeV resonance as an indication of $SU(3)_c\times SU(3)_L\times U(1)_X$ gauge symmetry,''
  arXiv:1512.07165 [hep-ph].
  %%CITATION = ARXIV:1512.07165;%%
  %28 citations counted in INSPIRE as of 04 Jan 2016
  %\cite{Dey:2015bur}
  \bibitem{Dey:2015bur}
  U.~K.~Dey, S.~Mohanty and G.~Tomar,
  %``750 GeV resonance in the Dark Left-Right Model,''
  arXiv:1512.07212 [hep-ph].
  %%CITATION = ARXIV:1512.07212;%%
  %29 citations counted in INSPIRE as of 04 Jan 2016
  %\cite{Pelaggi:2015knk}
  \bibitem{Pelaggi:2015knk}
  G.~M.~Pelaggi, A.~Strumia and E.~Vigiani,
  %``Trinification can explain the di-photon and di-boson LHC anomalies,''
  arXiv:1512.07225 [hep-ph].
  %%CITATION = ARXIV:1512.07225;%%
  %25 citations counted in INSPIRE as of 04 Jan 2016
  %\cite{Belyaev:2015hgo}
  \bibitem{Belyaev:2015hgo}
  A.~Belyaev, G.~Cacciapaglia, H.~Cai, T.~Flacke, A.~Parolini and H.~Serôdio,
  %``Singlets in Composite Higgs Models in light of the LHC di-photon searches,''
  arXiv:1512.07242 [hep-ph].
  %%CITATION = ARXIV:1512.07242;%%
  %26 citations counted in INSPIRE as of 04 Jan 2016
  %\cite{Huang:2015rkj}
  \bibitem{Huang:2015rkj}
  W.~C.~Huang, Y.~L.~S.~Tsai and T.~C.~Yuan,
  %``Gauged Two Higgs Doublet Model confronts the LHC 750 GeV di-photon anomaly,''
  arXiv:1512.07268 [hep-ph].
  %%CITATION = ARXIV:1512.07268;%%
  %22 citations counted in INSPIRE as of 04 Jan 2016
  %\cite{Cao:2015xjz}
  \bibitem{Cao:2015xjz}
  Q.~H.~Cao, S.~L.~Chen and P.~H.~Gu,
  %``Strong CP Problem, Neutrino Masses and the 750 GeV Diphoton Resonance,''
  arXiv:1512.07541 [hep-ph].
  %%CITATION = ARXIV:1512.07541;%%
  %20 citations counted in INSPIRE as of 04 Jan 2016
  %\cite{Gu:2015lxj}
  \bibitem{Gu:2015lxj}
  J.~Gu and Z.~Liu,
  %``Running after Diphoton,''
  arXiv:1512.07624 [hep-ph].
  %%CITATION = ARXIV:1512.07624;%%
  %17 citations counted in INSPIRE as of 04 Jan 2016
  %\cite{Patel:2015ulo}
  \bibitem{Patel:2015ulo}
  K.~M.~Patel and P.~Sharma,
  %``Interpreting 750 GeV diphoton excess in SU(5) grand unified theory,''
  arXiv:1512.07468 [hep-ph].
  %%CITATION = ARXIV:1512.07468;%%
  %19 citations counted in INSPIRE as of 04 Jan 2016
  %\cite{Badziak:2015zez}
  \bibitem{Badziak:2015zez}
  M.~Badziak,
  %``Interpreting the 750 GeV diphoton excess in minimal extensions of Two-Higgs-Doublet models,''
  arXiv:1512.07497 [hep-ph].
  %%CITATION = ARXIV:1512.07497;%%
  %18 citations counted in INSPIRE as of 04 Jan 2016
  %\cite{Chakraborty:2015gyj}
  \bibitem{Chakraborty:2015gyj}
  S.~Chakraborty, A.~Chakraborty and S.~Raychaudhuri,
  %``Diphoton resonance at 750 GeV in the broken MRSSM,''
  arXiv:1512.07527 [hep-ph].
  %%CITATION = ARXIV:1512.07527;%%
  %23 citations counted in INSPIRE as of 04 Jan 2016
  %\cite{Altmannshofer:2015xfo}
  \bibitem{Altmannshofer:2015xfo}
  W.~Altmannshofer, J.~Galloway, S.~Gori, A.~L.~Kagan, A.~Martin and J.~Zupan,
  %``On the 750 GeV di-photon excess,''
  arXiv:1512.07616 [hep-ph].
  %%CITATION = ARXIV:1512.07616;%%
  %23 citations counted in INSPIRE as of 04 Jan 2016
  %\cite{Cvetic:2015vit}
  \bibitem{Cvetic:2015vit}
  M.~Cvetic, J.~Halverson and P.~Langacker,
  %``String Consistency, Heavy Exotics, and the $750$ GeV Diphoton Excess at the LHC,''
  arXiv:1512.07622 [hep-ph].
  %%CITATION = ARXIV:1512.07622;%%
  %22 citations counted in INSPIRE as of 04 Jan 2016
  %\cite{Allanach:2015ixl}
  \bibitem{Allanach:2015ixl}
  B.~C.~Allanach, P.~S.~B.~Dev, S.~A.~Renner and K.~Sakurai,
  %``Di-photon Excess Explained by a Resonant Sneutrino in R-parity Violating Supersymmetry,''
  arXiv:1512.07645 [hep-ph].
  %%CITATION = ARXIV:1512.07645;%%
  %17 citations counted in INSPIRE as of 04 Jan 2016
  %\cite{Davoudiasl:2015cuo}
  \bibitem{Davoudiasl:2015cuo}
  H.~Davoudiasl and C.~Zhang,
  %``A 750 GeV Messenger of Dark Conformal Symmetry Breaking,''
  arXiv:1512.07672 [hep-ph].
  %%CITATION = ARXIV:1512.07672;%%
  %19 citations counted in INSPIRE as of 04 Jan 2016
  %\cite{Das:2015enc}
  \bibitem{Das:2015enc}
  K.~Das and S.~K.~Rai,
  %``The 750 GeV Diphoton excess in a $U(1)$ hidden symmetry model,''
  arXiv:1512.07789 [hep-ph].
  %%CITATION = ARXIV:1512.07789;%%
  %19 citations counted in INSPIRE as of 04 Jan 2016
  %\cite{Cheung:2015cug}
  \bibitem{Cheung:2015cug}
  K.~Cheung, P.~Ko, J.~S.~Lee, J.~Park and P.~Y.~Tseng,
  %``A Higgcision study on the 750 GeV Di-photon Resonance and 125 GeV SM Higgs boson with the Higgs-Singlet Mixing,''
  arXiv:1512.07853 [hep-ph].
  %%CITATION = ARXIV:1512.07853;%%
  %18 citations counted in INSPIRE as of 04 Jan 2016
  
  %\cite{Craig:2015lra}
  \bibitem{Craig:2015lra}
  N.~Craig, P.~Draper, C.~Kilic and S.~Thomas,
  %``How the $\gamma \gamma$ Resonance Stole Christmas,''
  arXiv:1512.07733 [hep-ph].
  %%CITATION = ARXIV:1512.07733;%%
  %14 citations counted in INSPIRE as of 04 Jan 2016
  %\cite{Liu:2015yec}
  \bibitem{Liu:2015yec}
  J.~Liu, X.~P.~Wang and W.~Xue,
  %``LHC diphoton excess from colorful resonances,''
  arXiv:1512.07885 [hep-ph].
  %%CITATION = ARXIV:1512.07885;%%
  %18 citations counted in INSPIRE as of 04 Jan 2016
  %\cite{Zhang:2015uuo}
  \bibitem{Zhang:2015uuo}
  J.~Zhang and S.~Zhou,
  %``Electroweak Vacuum Stability and Diphoton Excess at 750 GeV,''
  arXiv:1512.07889 [hep-ph].
  %%CITATION = ARXIV:1512.07889;%%
  %17 citations counted in INSPIRE as of 04 Jan 2016
  %\cite{Casas:2015blx}
  \bibitem{Casas:2015blx}
  J.~A.~Casas, J.~R.~Espinosa and J.~M.~Moreno,
  %``The 750 GeV Diphoton Excess as a First Light on Supersymmetry Breaking,''
  arXiv:1512.07895 [hep-ph].
  %%CITATION = ARXIV:1512.07895;%%
  %16 citations counted in INSPIRE as of 04 Jan 2016
  %\cite{Hall:2015xds}
  \bibitem{Hall:2015xds}
  L.~J.~Hall, K.~Harigaya and Y.~Nomura,
  %``750 GeV Diphotons: Implications for Supersymmetric Unification,''
  arXiv:1512.07904 [hep-ph].
  %%CITATION = ARXIV:1512.07904;%%
  %17 citations counted in INSPIRE as of 04 Jan 2016
  %\cite{Park:2015ysf}
  \bibitem{Park:2015ysf}
  J.~C.~Park and S.~C.~Park,
  %``Indirect signature of dark matter with the diphoton resonance at 750 GeV,''
  arXiv:1512.08117 [hep-ph].
  %%CITATION = ARXIV:1512.08117;%%
  %4 citations counted in INSPIRE as of 04 Jan 2016
  %\cite{Salvio:2015jgu}
  \bibitem{Salvio:2015jgu}
  A.~Salvio and A.~Mazumdar,
  %``Higgs Stability and the 750 GeV Diphoton Excess,''
  arXiv:1512.08184 [hep-ph].
  %%CITATION = ARXIV:1512.08184;%%
  %6 citations counted in INSPIRE as of 04 Jan 2016
  %\cite{Li:2015jwd}
  \bibitem{Li:2015jwd}
  G.~Li, Y.~n.~Mao, Y.~L.~Tang, C.~Zhang, Y.~Zhou and S.~h.~Zhu,
  %``A Loop-philic Pseudoscalar,''
  arXiv:1512.08255 [hep-ph].
  %%CITATION = ARXIV:1512.08255;%%
  %5 citations counted in INSPIRE as of 04 Jan 2016
  %\cite{Son:2015vfl}
  \bibitem{Son:2015vfl}
  M.~Son and A.~Urbano,
  %``A new scalar resonance at 750 GeV: Towards a proof of concept in favor of strongly interacting theories,''
  arXiv:1512.08307 [hep-ph].
  %%CITATION = ARXIV:1512.08307;%%
  %5 citations counted in INSPIRE as of 04 Jan 2016
  %\cite{An:2015cgp}
  \bibitem{An:2015cgp}
  H.~An, C.~Cheung and Y.~Zhang,
  %``Broad Diphotons from Narrow States,''
  arXiv:1512.08378 [hep-ph].
  %%CITATION = ARXIV:1512.08378;%%
  %4 citations counted in INSPIRE as of 04 Jan 2016
  %\cite{Wang:2015omi}
  \bibitem{Wang:2015omi}
  F.~Wang, W.~Wang, L.~Wu, J.~M.~Yang and M.~Zhang,
  %``Interpreting 750 GeV Diphoton Resonance in the NMSSM with Vector-like Particles,''
  arXiv:1512.08434 [hep-ph].
  %%CITATION = ARXIV:1512.08434;%%
  %4 citations counted in INSPIRE as of 04 Jan 2016
  %\cite{Cao:2015scs}
  \bibitem{Cao:2015scs}
  Q.~H.~Cao, Y.~Liu, K.~P.~Xie, B.~Yan and D.~M.~Zhang,
  %``The Diphoton Excess, Low Energy Theorem and the 331 Model,''
  arXiv:1512.08441 [hep-ph].
  %%CITATION = ARXIV:1512.08441;%%
  %7 citations counted in INSPIRE as of 04 Jan 2016
  %\cite{Gao:2015igz}
  \bibitem{Gao:2015igz}
  J.~Gao, H.~Zhang and H.~X.~Zhu,
  %``Diphoton excess at 750 GeV: gluon-gluon fusion or quark-antiquark annihilation?,''
  arXiv:1512.08478 [hep-ph].
  %%CITATION = ARXIV:1512.08478;%%
  %3 citations counted in INSPIRE as of 04 Jan 2016
  %\cite{Goertz:2015nkp}
  \bibitem{Goertz:2015nkp}
  F.~Goertz, J.~F.~Kamenik, A.~Katz and M.~Nardecchia,
  %``Indirect Constraints on the Scalar Di-Photon Resonance at the LHC,''
  arXiv:1512.08500 [hep-ph].
  %%CITATION = ARXIV:1512.08500;%%
  %3 citations counted in INSPIRE as of 04 Jan 2016
  %\cite{Dev:2015vjd}
  \bibitem{Dev:2015vjd}
  P.~S.~B.~Dev, R.~N.~Mohapatra and Y.~Zhang,
  %``Quark Seesaw Vectorlike Fermions and Diphoton Excess,''
  arXiv:1512.08507 [hep-ph].
  %%CITATION = ARXIV:1512.08507;%%
  %5 citations counted in INSPIRE as of 04 Jan 2016
  %\cite{Cao:2015apa}
  \bibitem{Cao:2015apa}
  J.~Cao, F.~Wang and Y.~Zhang,
  %``Interpreting The 750 GeV Diphton Excess Within TopFlavor Seesaw Model,''
  arXiv:1512.08392 [hep-ph].
  %%CITATION = ARXIV:1512.08392;%%
  %4 citations counted in INSPIRE as of 04 Jan 2016
  %\cite{Cai:2015hzc}
  \bibitem{Cai:2015hzc}
  C.~Cai, Z.~H.~Yu and H.~H.~Zhang,
  %``The 750 GeV diphoton resonance as a singlet scalar in an extra dimensional model,''
  arXiv:1512.08440 [hep-ph].
  %%CITATION = ARXIV:1512.08440;%%
  %3 citations counted in INSPIRE as of 04 Jan 2016
  %\cite{Kim:2015xyn}
  \bibitem{Kim:2015xyn}
  J.~E.~Kim,
  %``Is an axizilla possible for di-photon resonance?,''
  arXiv:1512.08467 [hep-ph].
  %%CITATION = ARXIV:1512.08467;%%
  %3 citations counted in INSPIRE as of 04 Jan 2016
  %\cite{Chao:2015nac}
  \bibitem{Chao:2015nac}
  W.~Chao,
  %``Neutrino Catalyzed Diphoton Excess,''
  arXiv:1512.08484 [hep-ph].
  %%CITATION = ARXIV:1512.08484;%%
  %3 citations counted in INSPIRE as of 04 Jan 2016
  %\cite{Anchordoqui:2015jxc}
  \bibitem{Anchordoqui:2015jxc}
  L.~A.~Anchordoqui, I.~Antoniadis, H.~Goldberg, X.~Huang, D.~Lust and T.~R.~Taylor,
  %``750 GeV diphotons from closed string states,''
  arXiv:1512.08502 [hep-ph].
  %%CITATION = ARXIV:1512.08502;%%
  %4 citations counted in INSPIRE as of 04 Jan 2016
  %\cite{Bizot:2015qqo}
  \bibitem{Bizot:2015qqo}
  N.~Bizot, S.~Davidson, M.~Frigerio and J.-L.~Kneur,
  %``Two Higgs doublets to explain the excesses $pp\rightarrow \gamma\gamma(750\ {\rm GeV})$ and $h \to \tau^\pm \mu^\mp$,''
  arXiv:1512.08508 [hep-ph].
  %%CITATION = ARXIV:1512.08508;%%
  %5 citations counted in INSPIRE as of 04 Jan 2016
  %\cite{Ibanez:2015uok}
  \bibitem{Ibanez:2015uok}
  L.~E.~Ibanez and V.~Martin-Lozano,
  %``A Megaxion at 750 GeV as a First Hint of Low Scale String Theory,''
  arXiv:1512.08777 [hep-ph].
  %%CITATION = ARXIV:1512.08777;%%
  %\cite{Huang:2015svl}
  \bibitem{Huang:2015svl}
  X.~J.~Huang, W.~H.~Zhang and Y.~F.~Zhou,
  %``A 750 GeV dark matter messenger at the Galactic Center,''
  arXiv:1512.08992 [hep-ph].
  %%CITATION = ARXIV:1512.08992;%%
  %\cite{Chiang:2015tqz}
  \bibitem{Chiang:2015tqz}
  C.~W.~Chiang, M.~Ibe and T.~T.~Yanagida,
  %``Revisiting Scalar Quark Hidden Sector in Light of 750-GeV Diphoton Resonance,''
  arXiv:1512.08895 [hep-ph].
  %%CITATION = ARXIV:1512.08895;%%
  %\cite{Kang:2015roj}
  \bibitem{Kang:2015roj}
  S.~K.~Kang and J.~Song,
  %``Top-phobic heavy Higgs boson as the 750 GeV diphoton resonance,''
  arXiv:1512.08963 [hep-ph].
  %%CITATION = ARXIV:1512.08963;%%
  
  %\cite{Kanemura:2015bli}
  \bibitem{Kanemura:2015bli}
  S.~Kanemura, K.~Nishiwaki, H.~Okada, Y.~Orikasa, S.~C.~Park and R.~Watanabe,
  %``LHC 750 GeV Diphoton excess in a radiative seesaw model,''
  arXiv:1512.09048 [hep-ph].
  %%CITATION = ARXIV:1512.09048;%%
  %\cite{Low:2015qho}
  \bibitem{Low:2015qho}
  I.~Low and J.~Lykken,
  %``Implications of Gauge Invariance on a Heavy Diphoton Resonance,''
  arXiv:1512.09089 [hep-ph].
  %%CITATION = ARXIV:1512.09089;%%
  %\cite{Hernandez:2015hrt}
  \bibitem{Hernandez:2015hrt}
  A.~E.~C.~Hernández,
  %``The 750 GeV diphoton resonance can cause the SM fermion mass and mixing pattern,''
  arXiv:1512.09092 [hep-ph].
  %%CITATION = ARXIV:1512.09092;%%
  %\cite{Kaneta:2015qpf}
  \bibitem{Kaneta:2015qpf}
  K.~Kaneta, S.~Kang and H.~S.~Lee,
  %``Diphoton excess at the LHC Run 2 and its implications for a new heavy gauge boson,''
  arXiv:1512.09129 [hep-ph].
  %%CITATION = ARXIV:1512.09129;%%
  %\cite{Dasgupta:2015pbr}
  \bibitem{Dasgupta:2015pbr}
  A.~Dasgupta, M.~Mitra and D.~Borah,
  %``Minimal Left-Right Symmetry Confronted with the 750 GeV Di-photon Excess at LHC,''
  arXiv:1512.09202 [hep-ph].
  %%CITATION = ARXIV:1512.09202;%%
  
  
  
  

  
     
    %\cite{King:2005jy}
\bibitem{King:2005jy}
  S.~F.~King, S.~Moretti and R.~Nevzorov,
  %``Theory and phenomenology of an exceptional supersymmetric standard model,''
  Phys.\ Rev.\ D {\bf 73} (2006) 035009
  doi:10.1103/PhysRevD.73.035009
  [hep-ph/0510419];
  %%CITATION = doi:10.1103/PhysRevD.73.035009;%%
  %163 citations counted in INSPIRE as of 25 Dec 2015
  %\cite{King:2005my}
%\bibitem{King:2005my}
  S.~F.~King, S.~Moretti and R.~Nevzorov,
  %``Exceptional supersymmetric standard model,''
  Phys.\ Lett.\ B {\bf 634} (2006) 278
  doi:10.1016/j.physletb.2005.12.070
  [hep-ph/0511256].
  %%CITATION = doi:10.1016/j.physletb.2005.12.070;%%
  %101 citations counted in INSPIRE as of 25 Dec 2015
  
  %\cite{Howl:2007zi}
\bibitem{Howl:2007zi}
  R.~Howl and S.~F.~King,
  %``Minimal E(6) Supersymmetric Standard Model,''
  JHEP {\bf 0801} (2008) 030
  doi:10.1088/1126-6708/2008/01/030
  [arXiv:0708.1451 [hep-ph]].
  %%CITATION = doi:10.1088/1126-6708/2008/01/030;%%
  %45 citations counted in INSPIRE as of 25 Dec 2015
      
   
%\cite{Callaghan:2011jj}   
\bibitem{Callaghan:2011jj}
  J.~C.~Callaghan, S.~F.~King, G.~K.~Leontaris and G.~G.~Ross,
  %``Towards a Realistic F-theory GUT,''
  JHEP {\bf 1204} (2012) 094
  doi:10.1007/JHEP04(2012)094
  [arXiv:1109.1399 [hep-ph]].
  %%CITATION = doi:10.1007/JHEP04(2012)094;%%
  %22 citations counted in INSPIRE as of 25 Dec 2015

%\cite{Callaghan:2012rv}
\bibitem{Callaghan:2012rv}
  J.~C.~Callaghan and S.~F.~King,
  %``E6 Models from F-theory,''
  JHEP {\bf 1304} (2013) 034
  doi:10.1007/JHEP04(2013)034
  [arXiv:1210.6913 [hep-ph]].
  %%CITATION = doi:10.1007/JHEP04(2013)034;%%
  %15 citations counted in INSPIRE as of 25 Dec 2015

%\cite{Callaghan:2013kaa}
\bibitem{Callaghan:2013kaa}
  J.~C.~Callaghan, S.~F.~King and G.~K.~Leontaris,
  %``Gauge coupling unification in $E_6$ F-theory GUTs with matter and bulk exotics from flux breaking,''
  JHEP {\bf 1312} (2013) 037
  doi:10.1007/JHEP12(2013)037
  [arXiv:1307.4593 [hep-ph]].
  %%CITATION = doi:10.1007/JHEP12(2013)037;%%
  %7 citations counted in INSPIRE as of 25 Dec 2015
  
 %\cite{Hall:2012mx}
\bibitem{Hall:2012mx}
  J.~P.~Hall and S.~F.~King,
  %``Nmssm+,''
  JHEP {\bf 1301} (2013) 076
  doi:10.1007/JHEP01(2013)076
  [arXiv:1209.4657 [hep-ph]].
  %%CITATION = doi:10.1007/JHEP01(2013)076;%%
  %14 citations counted in INSPIRE as of 02 Jan 2016
   
   %\cite{Djouadi:2005gi}
\bibitem{Djouadi:2005gi}
  A.~Djouadi,
  %``The Anatomy of electro-weak symmetry breaking. I: The Higgs boson in the standard model,''
  Phys.\ Rept.\  {\bf 457} (2008) 1
  doi:10.1016/j.physrep.2007.10.004
  [hep-ph/0503172].
  %%CITATION = doi:10.1016/j.physrep.2007.10.004;%%
  %998 citations counted in INSPIRE as of 04 Jan 2016


\end{thebibliography}
\end{document}